\documentclass[useAMS,usenatbib]{mn2e}
\usepackage{graphicx}
\usepackage{natbib}  
\usepackage{epstopdf}
\newcommand\aj{{AJ}}%
\newcommand\apj{{ApJ}}%
\newcommand\apjl{{ApJ}}%
\newcommand\aap{{A\&A}}%
\newcommand\mnras{{MNRAS}}%
\newcommand\nat{{Nature}}%
\def\f435{$m_{\rm F435W}$}
\def\ff625{$m_{\rm F625W}$}
\def\simgt{\lower.5ex\hbox{$\; \buildrel > \over \sim \;$}}
\def\simlt{\lower.5ex\hbox{$\; \buildrel < \over \sim \;$}}

\newcommand\teff{T$_{\rm eff}$}

\newcommand\Teff{T$_{\rm eff}$}

\newcommand\Mc{M$_{\rm core}$}
\newcommand\mod{$m_{\rm F435W}-M_{\rm F435W}$}

\newcommand{\msun}{\ensuremath{\, {M}_\odot}}
\newcommand{\Msun}{\ensuremath{\, {M}_\odot}}
\newcommand{\ocen}{$\omega$~Cen}

\title[The evolutionary status of the blue hook stars in \ocen]{The evolutionary status of the blue hook stars in $\omega$ Centauri}
\author[F. D'Antona, V. Caloi, \& P. Ventura]{Francesca D'Antona$^{1}$, Vittoria Caloi$^{2}$ and
Paolo Ventura$^{1}$
\thanks{E-mail: dantona@oa-roma.inaf.it (FD); 
vittoria.caloi@iasf-roma.inaf.it (VC); ventura@oa-roma.inaf.it (PV)}
\footnotemark[1]
\\
$^{1}$INAF, Osservatorio Astronomico di Roma, Via Frascati 33, 00040 Monteporzio Catone (Roma), Italy.\\
$^{2}$ INAF, IASF--Roma, via Fosso del Cavaliere 100, I-00133 Roma, Italy\\
}
\begin{document}

\date{Accepted . Received ; in original form }

\pagerange{\pageref{firstpage}--\pageref{lastpage}} \pubyear{2006}

\maketitle

\label{firstpage}

\begin{abstract}
Core helium burning is the dominant source of energy of extreme horizontal branch stars, 
as the hydrogen envelope is too small to contribute to the nuclear energy output. The evolution 
of each mass in the HR diagram occurs along vertical tracks that, when the core helium
is consumed, evolve to higher \teff\ and then to the white dwarf stage. The larger is the mass, the
smaller is the \teff\ of the models, so that the zero age horizontal branch (ZAHB) is ``horizontal". 
In this paper we show that, if the helium mass fraction (Y) of the envelope is larger 
than Y$\sim$0.5, the shape of the tracks changes completely: the hydrogen burning 
becomes efficient again also for very small envelope masses, thanks
to the higher molecular weight and to the higher temperatures of the hydrogen shell. The larger is Y,
the smaller is the envelope mass that provides strong H--shell burning. These tracks have a curled shape,
are located at a \teff\ following the approximate relation \teff=8090+ 32900$\times$Y, 
and become more luminous for larger envelope masses. Consequently, the ZAHB of the very 
high helium models is ``vertical" in the HR diagram. 
Synthetic models based on these tracks nicely reproduce the location and shape of the
``blue hook" in the globular cluster \ocen, best fit by a very high \teff\ (bluer) sequence  
with Y=0.80 and a cooler (redder) one with Y=0.65. Although these precise values of Y may depend 
on the color--\teff\ conversions, we know that the helium content of
the progenitors of the blue hook stars can not be larger than Y$\sim$0.38--0.40,
if they are descendants of the cluster blue main sequence. Consequently, 
this interpretation implies that
all these objects must in fact be progeny of the blue main sequence, but they have 
all suffered further deep mixing, that has largely and uniformly 
increased their surface helium abundance, during the red giant branch evolution. 
A late helium flash can not be the cause of this deep mixing, as 
the models we propose have hydrogen rich envelopes much more massive than those
required for a late flash. We discuss different models of deep mixing proposed in the literature,
and conclude that our interpretation of the blue hook can not be ruled out, 
but requires a much deeper investigation before it can be accepted.
\end{abstract}

\begin{keywords}
globular clusters; chemical abundances; self-enrichment
\end{keywords}

\section{Introduction}
\label{sec:intro}
The vast problematics involving the globular cluster $\omega$ Cen goes from
  the presence of multiple evolutionary sequences all along the main sequence
  \citep{bedin2004},
  the subgiant and the giant branches \citep{lee1999nature,pancino2000,rey2004,sollima2005}, 
  to the presence of a wide spread in metal content (e.g., Norris \& Da Costa 1995; Suntzeff \& Kraft
1996; Smith et al. 2000; Johnson et al. 2009) and a likely spread in helium content
\citep{norris2004,piotto2005}. Even if present in
  other clusters, these features are not all found together as in this most
  massive GC in the Galaxy. Another striking feature in the colour-magnitude
  (CM) diagram of $\omega$ Cen is the large number of very faint and very blue
  stars at the end of the horizontal branch (HB), beyond the temperature and
  magnitude limits attainable by currently known HB structures. Such objects
  are commonly called "blue hook" stars and are found in other clusters ---M54 \citep{rosenberg2004},
  NGC 2808 \citep{brown2001}, NGC 2419 \citep{ripepi2007}, perhaps NGC 6388 \citep{busso2007}--- 
  but in $\omega$ Cen their
  distribution in the CM diagram is different from what observed elsewhere
  (Figure \ref{f2}, left panel). Most of the blue hook objects are arranged
  along two ``parallel sequences'' inclined towards high temperatures \citep{anderson2009}. 
Spectroscopic analysis of the blue hook stars has been done by
\cite{moehler2004} for NGC~2808 resulting in \teff$\sim$35--36$\times 10^3$K for the hottest
objects, with a couple of stars probably already evolving towards the white dwarf stage. In
\ocen, they are in the range 32--36$\times 10^3$K, according to \cite{moehler2007}, but many hotter
stars are also present. More recently, \cite{villanova2009} have re--examined several blue hook stars in \ocen\ 
and find a clustering in a range of \teff$\sim 35-38 \times 10^3$K. 
Being the spectroscopic tools hampered 
by uncertainties much larger than the HST photometry, we can infer from the vertical shape of the
extreme blue hook that all its stars cluster at very similar \teff. 
We do know that these extreme HB stars should be mainly burning helium in their cores, and evolve 
along a ``vertical" track that can simulate the shape of the blue hook, but without covering 
its whole luminosity extension. In fact, in order to explain the blue hook with tracks of
this kind, \cite{cassisi2009} adopt a superposition of stars having a normal
helium content (that start their evolution at a larger luminosity) plus stars with enhanced helium
(starting their evolution at lower luminosity), 
born from the high helium blue MS stars, plus stars that suffered a late helium core flash, and are
then also carbon enhanced \citep{moehler2007}. 
A reason to justify the hypothesis that the blue hook is the superposition of the progeny of very different 
evolutionary parents comes from
the analysis of helium abundance, since blue hook stars show a great variety of results: they are 
both helium--normal and helium rich objects, the latter ones 
close to the range Y=0.38--0.40 believed to apply to
the blue MS \citep[see, e.g. figure 2 in][]{moehler2007}. 
The hottest (\teff$>$40000K) stars are generally extremely helium rich, or even very helium poor.
In our opinion, however, it is not necessary to connect these different helium abundances to different 
evolutionary paths. The atmospheric analysis is very difficult, and the resulting surface 
helium abundance may be affected by errors difficult to be fully quantified. In addition,
it is well possible that the main reasons why the surface abundances of
these stars are largely different from each other, are surface phenomena acting 
on timescales much shorter than the evolutionary time in HB, like helium sedimentation 
in the thin atmospheric layer, or residual mass loss. The presence of a well defined locus
in the HR diagram, together with what looks like a broader (cooler) sequence at redder colors 
(see Figure 2, left panel) appears more like some kind of evolutionary boundary
that limits the stars' evolution, than as the casual result of many different
evolutionary paths. The HST data set shown in Figure 2 is the same used and fully
described in \cite{cassisi2009}, 
namely a mosaic of 3$\times$3 fields obtained with the ACS/WFC (GO--9442, PI A. Cool)
through the F435W and F625W filters.  
More than 350 stars populate the thin vertical part of the blue hook,
and $\sim$170 are spread on the right of this sequence. 
\begin{figure*}
\includegraphics[width=8cm]{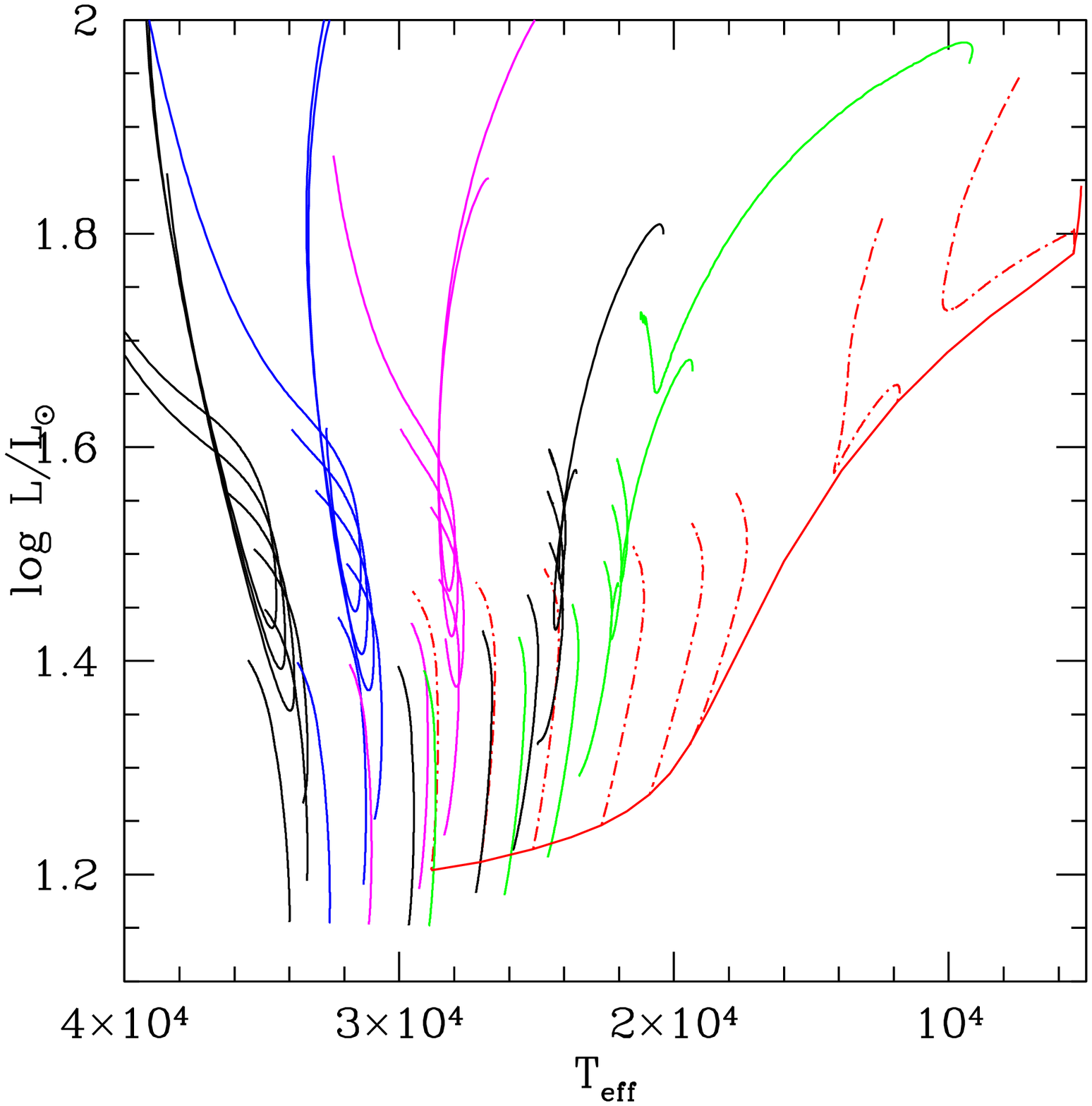}
\includegraphics[width=8cm]{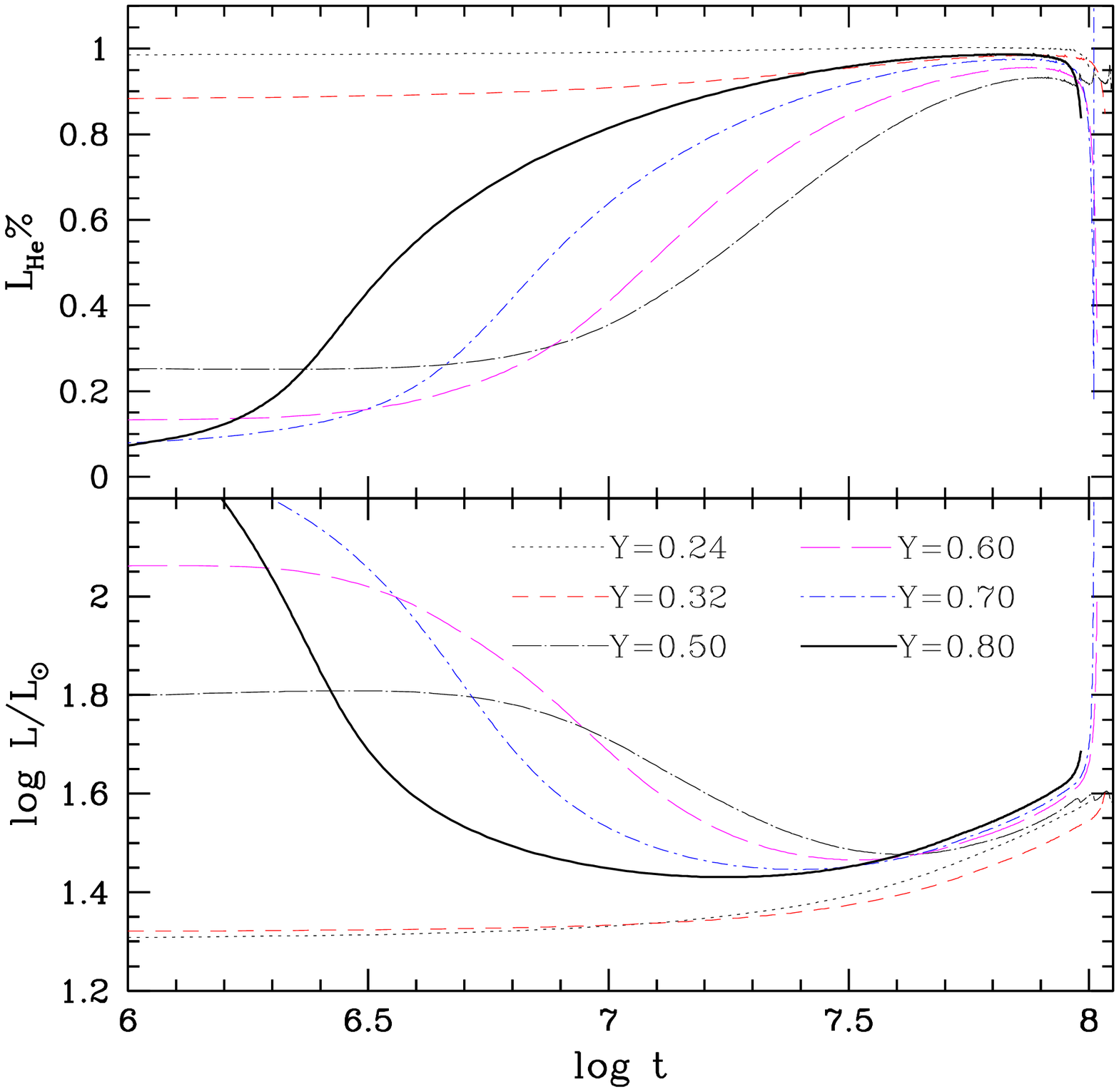}
\caption{On the left we plot the evolutionary tracks of different total masses for surface 
helium content Y=0.80, 0.70, 0.60, 0.50, 0.45
from left to right. Total masses for Y=0.8, 0.7 and 0.6 are 0.46, 0.47, 0.48, 0.49, 0.50 and 0.51\msun. 
For Y=0.45 we have added M=0.55\msun.
The core mass is fixed at \Mc=0.455\msun. 
For comparison, we add the ZAHB and the corresponding (dash--dotted) tracks, having Y=0.32, total 
masses 0.472, 0.475, 0.48, 0.49, 0.50, 0.51, 0.55 and 0.60\msun\ 
from left to right, and \Mc=0.469\msun: these tracks
do not show the curling shape at any mass. The transition from pure--core helium burning
tracks to tracks in which also hydrogen burning is efficient is at $\sim$0.53\msun.
On the right side we plot (top panel) the \% of luminosity provided by the core helium burning
for a mass of 0.51\msun, with Y=0.24, 0.32, 0.5, 0.60, 0.70 and 0.8. 
Decreasing Y, the hydrogen shell burning contributes a larger fraction of luminosity for
a longer time. If Y$<$0.5, however, the hydrogen shell does not ignite any longer for this 
mass. In the bottom panel, we see that, for Y$\geq$0.5, the models spend most of their lifetime
in the ``curling" region (see left panel and text).}
      \label{f1}%
\end{figure*}
\begin{figure}
\includegraphics[width=8cm]{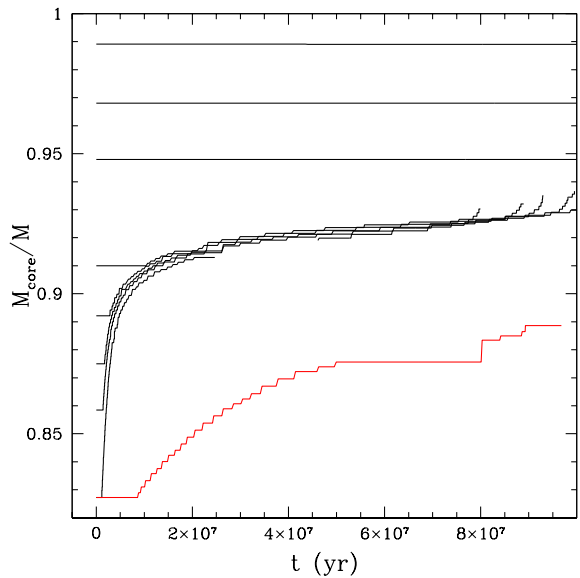}
\caption{We plot the ratio {\sl q} of the core mass to total mass as a function of time for the
Y=0.80 tracks of 0.46, 0.47, 0.48, 0.50, 0.51, 0.52, 0.53 and 0.55\Msun\ (from top to bottom). 
When the H--burning
begins to be efficient, {\sl q} rapidly increases due to the efficiency of the H--shell, giving
the same values of {\sl q} for all these masses. Thanks to
the same {\sl q}, the tracks remain at the same \teff, at
increasing luminosity, due to the larger total mass. The lowest track plotted
shows the {\sl q} for M=0.55\Msun\ and Y=0.45: the lower shell efficiency does not allow
the fast increase of the core mass.  
}
      \label{f1bis}%
\end{figure}

In Section 2 we present new models of helium core burning stars of metallicity adequate to describe 
\ocen's stars, characterized by extremely high values of helium abundance 
(from Y=0.45 to Y=0.8) around the burning helium core.
While, in Section 3, we will give reasonable but unproven justifications
for adopting these extravagant compositions, 
these models are the first that can reproduce the vertical shape of the blue hook stars
in \ocen, as the ZAHB of these models is
``vertical", and not ``horizontal" as it happens for less helium rich compositions.
Simulations of the blue hook star population allow us to speculate on the evolutionary 
path leading to this kind of stellar structures.

\section{The evolution of extremely helium rich HB stars}
The models presented in this work have been computed with the ATON code, by using the
recently updated input physics described in \cite{ventura2009}. We adopt
a mixture (Grevesse \& Sauval 1998) having [Fe/H]=--1.6, $\alpha$--enhanced, with $[\alpha$/Fe$]=0.4$,
for the latest opacities by Ferguson et al. (2005) 
at temperatures lower than 10000~K, and the OPAL opacities in the version 
documented by Iglesias \& Rogers (1996). 
Electron conduction opacities were  taken  from the WEB site of Potekhin 
(see the web page http://www.ioffe.rssi.ru/astro/conduct/ dated 2006) and
correspond to the \cite{potekhin1999} treatment. The electron opacities
are harmonically added to the radiative opacities.
In \cite{ventura2009} we employed three different mixtures for the C, N and O abundances, in
order to test the difference between ``standard" C+N+O and C+N+O enriched models. 
In this paper we use models having the standard CNO mixtures and the CNOx5 mixture (see later).
We compute HB models for different helium content Y. Models from Y=0.24 to Y=0.40
will be presented in another paper \citep{ventura2009b}, here we show the results for 
models with 0.45$\simlt$Y$\simlt$0.80. 
Generally, the HB models are computed starting from the zero age HB (ZAHB) by fixing the helium core mass.
In an evolutionary context, this core mass must be equal to the mass at the core helium flash
of the stars evolving on the red giant branch at the plausible age of the cluster. In our
standard models, we adopt 
the core mass resulting at ages in the range 10-12Gyr, for Y values from Y=0.24 to Y=0.40.
For the computations presented here, we make the hypothesis that values of Y$>$0.40 are
due to deep mixing, and are not the starting main sequence value.
Consequently, we adopt the core helium flash mass appropriate for the chosen chemistry and
for Y=0.40, that is M=0.455\msun\ for the CNOx5 composition (but see later for the size of the
helium core).
{\it We then are making the implicit hypothesis that our HB models progenitors are stars
belonging to the blue MS of \ocen} as will be justified in Sect. 3. 
\begin{figure*}
\includegraphics[width=5.5cm]{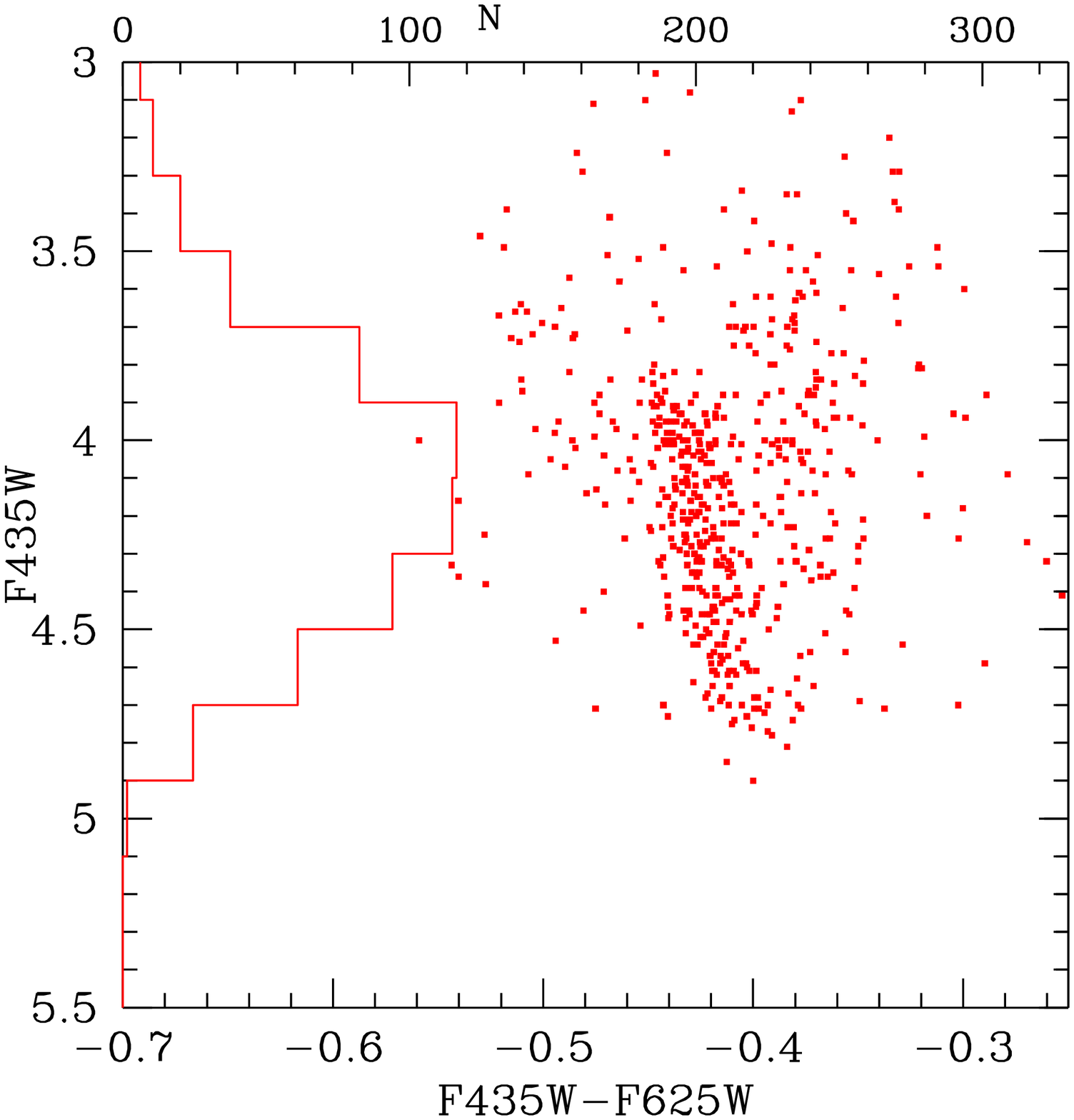}
\includegraphics[width=5.5cm]{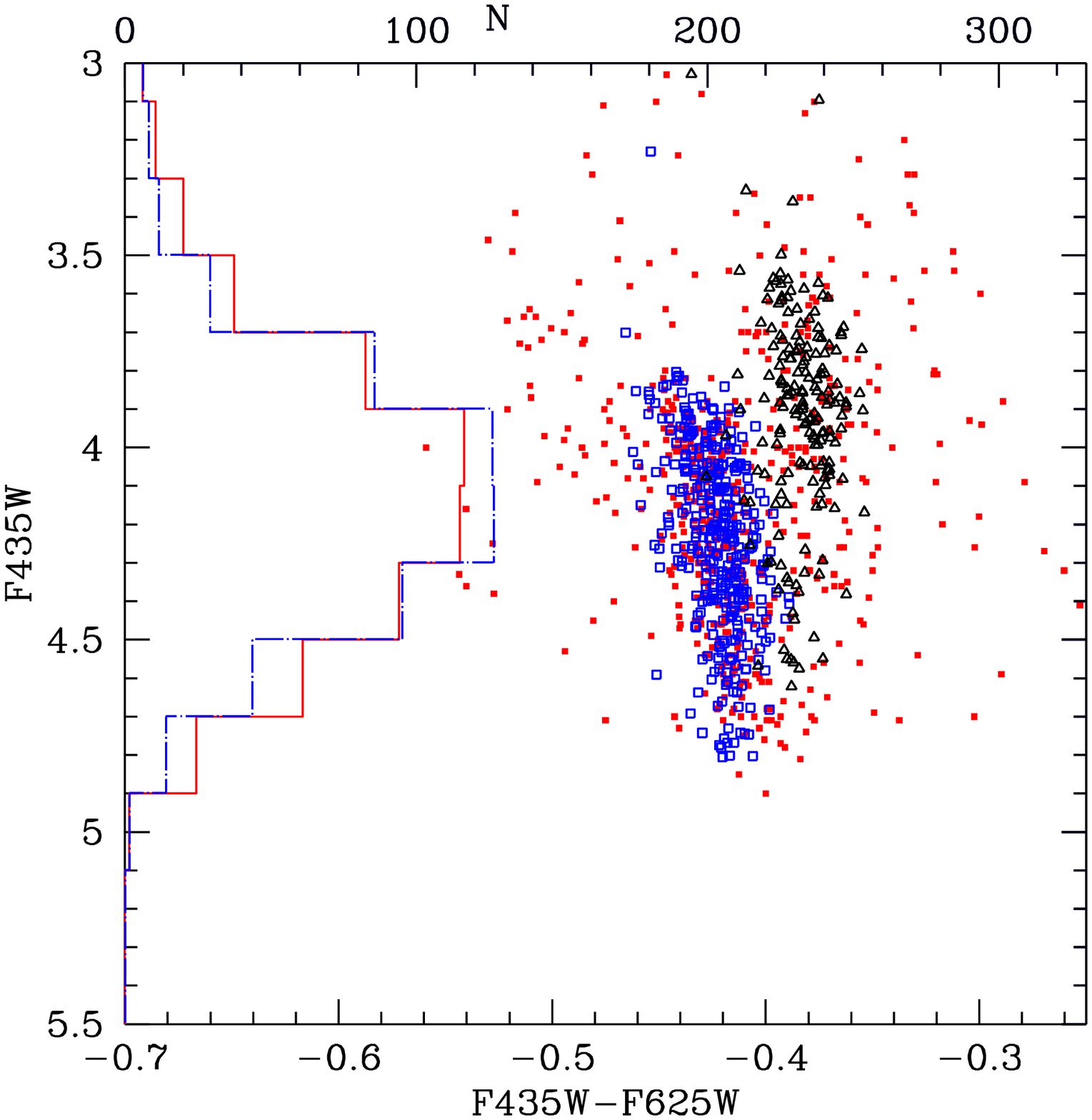}
\includegraphics[width=5.5cm]{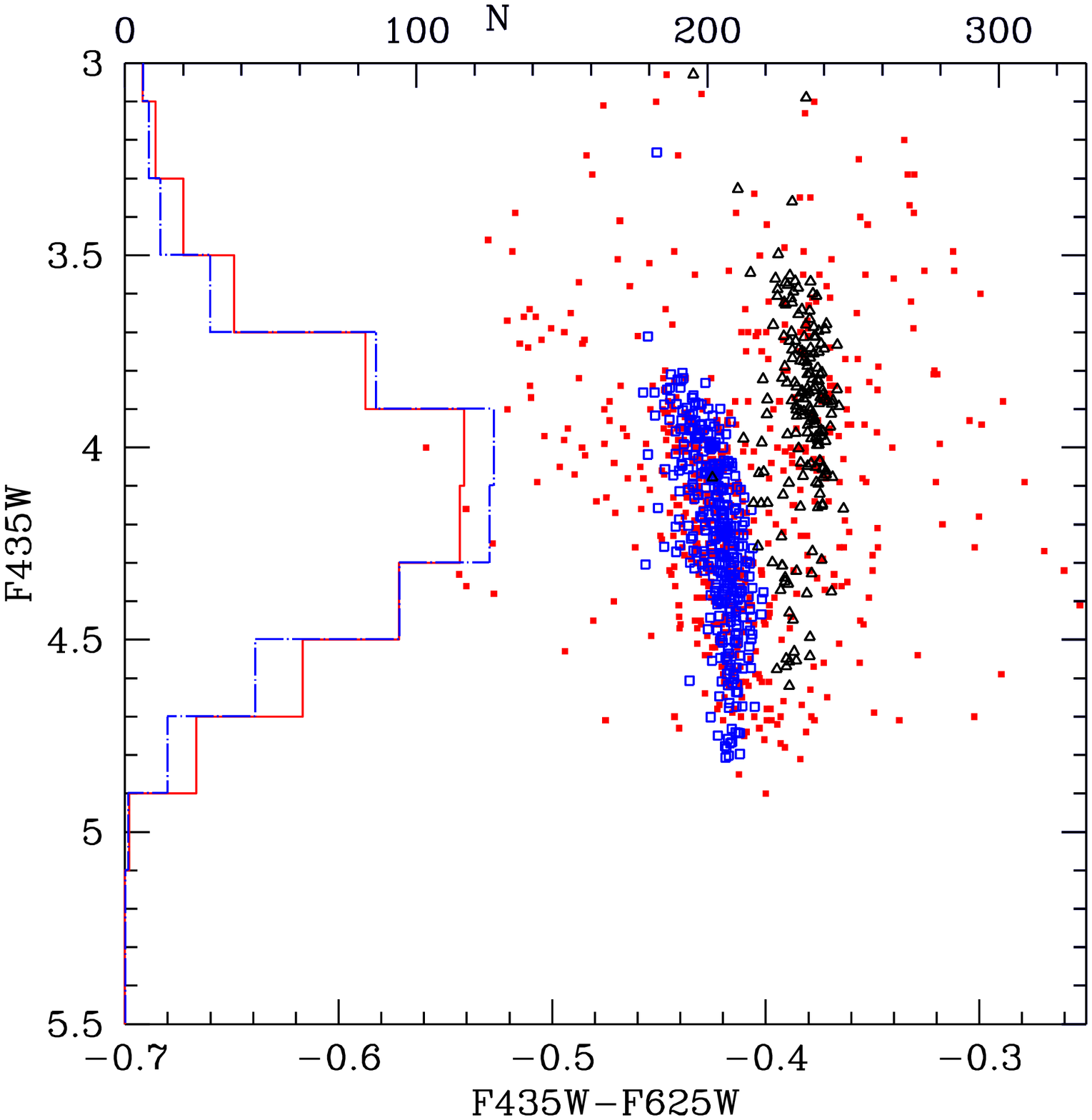}
\caption{On the left we plot
the blue hook in \ocen\ in the ACS/WFC mosaic reduced and described
in Anderson \& van der Marel (2009), for the magnitudes \f435\ and \ff625. 
The ACS/WFC mosaic covers about 10'$\times$10' centered on the cluster. 
Two separate, quasi--parallel ``vertical" sequences are evident.
The simulations of the blue hook are shown in the center and right side. In the center 
figure, we assume $\sigma$=0.01mag on both magnitudes, in
the right figure the $\sigma$ is reduced to 0.005mag. 
The full squares are the observed data points of Figure\ref{f1}.
Superimposed, the open (blue) squares represent the Y=0.80 sample, including a total of 350
extractions, while the open (black) triangles represent a sample of 170 stars having Y=0.65. Details are
given in the text. The histograms represent the observed sample (full line --red) and the total
simulation (dash--dotted line --blue). 
}
      \label{f2}%
\end{figure*}  

  In Figure \ref{f1} we show the evolution in the theoretical HR diagram
  for stars with Y=0.45, 0.50, 0.60, 0.70 and 0.80. Smaller values of Y do not show a ``blue hook type"
  behaviour, and we plot models for Y=0.32, CNOx1 (here the evolutionary core mass is 0.469\Msun),
  together with its ZAHB, to show its ``horizontal" behaviour.\\
Let us describe the Y=0.8 tracks. The smallest total masses show the
typical vertical track dominated by helium core burning: in fact, the hydrogen shell is not active, 
due to the very small hydrogen buffer on the core. 
As soon as the H--shell is ignited, its strength (due to the high molecular weigth) is such that
the ZAHB position jumps to very large luminosity. In $\sim$10Myr, the structure burns a large fraction of its hydrogen
envelope, with a strong decrease in luminosity. At this stage, He--burning provides more than $\sim$90\%
of the energy output, and the tracks revert their path increasing their  luminosity and giving rise
to the curling feature in Fig. \ref{f1}. A similar behaviour is found as long as  Y$\geq$0.5.

For all these Y's, the masses high enough to develop a H--shell are pushed into the same interval in \teff,
with the luminosity slightly increasing with the total mass. 
The reason for this occurrence is shown in Fig. \ref{f1bis}, where we plot the ratio of
core mass to total mass ({\sl q}) as a function of time. As long as the H--shell is not active, {\sl q}
does not increase, and the \Teff\ location decreases with decreasing  {\sl q}, as expected.
The appearance of an efficient H--shell gives origin to a rapid increase in  {\sl q}, such that
the values converge to an almost identical {\sl q} vs. time relation for all masses. Consequently, 
since \Teff\ depends on {\sl q}, the tracks cluster at similar \teff, with luminosity increasing 
with mass.  
  Figure \ref{f1} shows that the location of this curling--type evolution depends 
on Y: the larger is Y, the larger is the \teff. 
Approximately, we can use the linear relation $T_{\rm eff}=8090+ 32900\times Y$ to describe 
the \teff\ location of these tracks as a function of the helium content for $0.50\leq$Y$\leq$0.80.
This peculiar behaviour does not occurr at Y$\leq$0.45,
and this may be the reason why it has never been appreciated before.

Our choice of chemistry does not affect sensibly the \teff\ location of the tracks: models with
the same metallicity and normal CNO (the CNOx1 set) lie at about equal \teff, for the same choice of Y.
The luminosity location of the tracks however depends on the chosen core mass. 
We use the core mass corresponding to CNOx5 because, being smaller than for CNOx1, the fit 
with the observed luminosity of blue hook stars is easier, all other features being the 
same if we had chosen the core mass for CNOx1. We come back to this point when 
describing the simulations.

If we transform the tracks into the
HST filters of Figure \ref{f2}, we recognize that 
  the Y=0.80 sequence formed by HB structures with masses from 0.46  to 0.52\msun\
  overlaps nicely with the more populated and hotter vertical sequence in the blue hook in
  \ocen, even reproducing its slightly skew appearance. This is an important feature
  of present result, that to our knowledge had not been obtained by other
  evolutionary interpretations. The rest of the blue hook stars, slightly cooler,
  are located in between the Y=0.8 and Y=0.6 sequences.
  An attempt to simulate the blue hook is shown in Figure \ref{f2}. We extract a number 
  of stars with Y=0.80, plus a smaller sample with Y=0.65 (interpolated
  between the Y=0.6 and Y=0.7 sets of tracks). We choose an
  average mass of 0.50\msun, with a gaussian dispersion $\sigma$=0.023\msun.
  When the random mass extraction provides a value below 0.457\msun, we impose that this mass is
  not able to ignite helium and becomes a helium white dwarf. In this way, the random
  extractions produce a population of 95 helium WDs. The total number of extractions
  is such that we obtain in total the observed number of 350 stars for the
  bluest part of the hook (Y=0.80 in the simulation) and 170 stars for the redder side (Y=0.65
  in the simulation).
  A similar result is obtained if we make the hypothesis that the helium content in the
  redder stars varies randomly between 0.6 and 0.7
  (or even between 0.6 and 0.8). The results would not be equivalent if we could deal with
  a sample in which the observational errors are smaller. In Figure \ref{f2} the central 
  panel shows the simulation imposing an error of 0.01mag in the color, while the error is
  only 0.005mag in the right panel. If two separate sequences exist, they will be distinguished
  with smaller errors.
  The best representation of the data is obtained by assuming a distance modulus 
  \mod=14.45mag, while, e.g. \cite{cassisi2009} adopt 14.75mag. This modulus is the best one for 
  matching our models to the observed blue hook, but possible variations in the
  physical parameters of the models can end up with a different fit\footnote{
  By this choice of distance modulus, and keeping a mass loss during the giant evolution
  independent of the helium content, the bulk of luminous blue HB stars can be simulated with 
  Y=0.29, while the red (sparse) HB stars are compatible with Y=0.25. The HB would then find
  an interpretation similar to that given for NGC~2808 \citep{dantona2005}, but notice that
  metallicity differences should also be included in a consistent simulation.
  The choice of \cite{cassisi2009} modulus makes the luminous part of the HB very overluminous
  for our models.}. 
  As we have already pointed out, the distance modulus depends on the adopted helium
  core mass \Mc=0.455\msun. Larger \Mc's provide more luminous models, and the fit 
  requires a larger distance modulus, and viceversa for a smaller \Mc. 
  Also notice that we can expect smaller core masses
  when the evolving giant loses an amount of mass such as to reduce the envelope to M$_{\rm env} \sim
  0.003 - 0.008$\msun\ \citep{dcruz1996}. In this case, an early or late helium flash can follow
  (but this is not relevant in our modelling of the hook). 
  Further, if the blue hook is interpreted with stars having Y$>$0.60, and helium 
  enrichment takes place during the upper RGB evolution, then the
  core mass should be the one corresponding to the actual value of the
  envelope Y at the moment of the flash. This would imply a substantially smaller core. 
  
  A very critical parameter of the fit is the envelope helium abundance: we
  find that the models having Y=0.80 are appropriate, but the color-\teff\ transformations adopted 
  have been computed for a normal helium abundance, so that the required Y may be different (although in any case very
  large) if more appropriate transformations give a different value for the color  location of the
  vertical extreme HBs.  
  
  There are several appealing features in this simulation: 1) the vertical shape of the left
  blue hook is reproduced extremely well, {\it by using a unique set of evolutionary tracks} and
  not a superposition of different evolutions, whose contemporary occurrence must be finely tuned.
  2) the right side of the blue hook is also interpreted by the same kind of 
  ``vertical" evolution, possibly 
  with some spread in the helium abundance characteristic of the envelope of these stars, in contrast to
  the unique value required to reproduce the left side. 3) the luminosity function can also be
  reproduced, by adjusting the average value of the mass in the hook and the gaussian dispersion around
  this value. 
  
  We can try to simulate the blue hook by adopting tracks without the curling, by choosing, e.g., Y=0.45
  tracks. For the set adopted here, the core mass is \Mc=0.468\msun\ and Z=0.001 (solar scaled CNO). 
  The simulations are shown in Figure 3. If we adopt an average mass of 0.475\msun\
  of  and a mass dispersion of 0.01\msun\ or even as low as 0.004\msun, the shape of 
  the simulation is clearly very 
  different from the observed blue hook. 
  If we reduce the mass dispersion to 0.002\msun, the shape
  resembles the cooler component of the blue hook. For smaller core mass, we can model smaller
  total masses, that have bluer colors, but the shape of the simulation would be very similar
  and would not be consistent with the observed bluer hook. We conclude
  that the very high helium, curling tracks are particularly apt 
  in reproducing the blue hook.
  
  If we accept this interpretation, however, we have a very strong constraint of 
  the evolution leading to the blue hook phase.

\begin{figure*}
\includegraphics[width=4.cm]{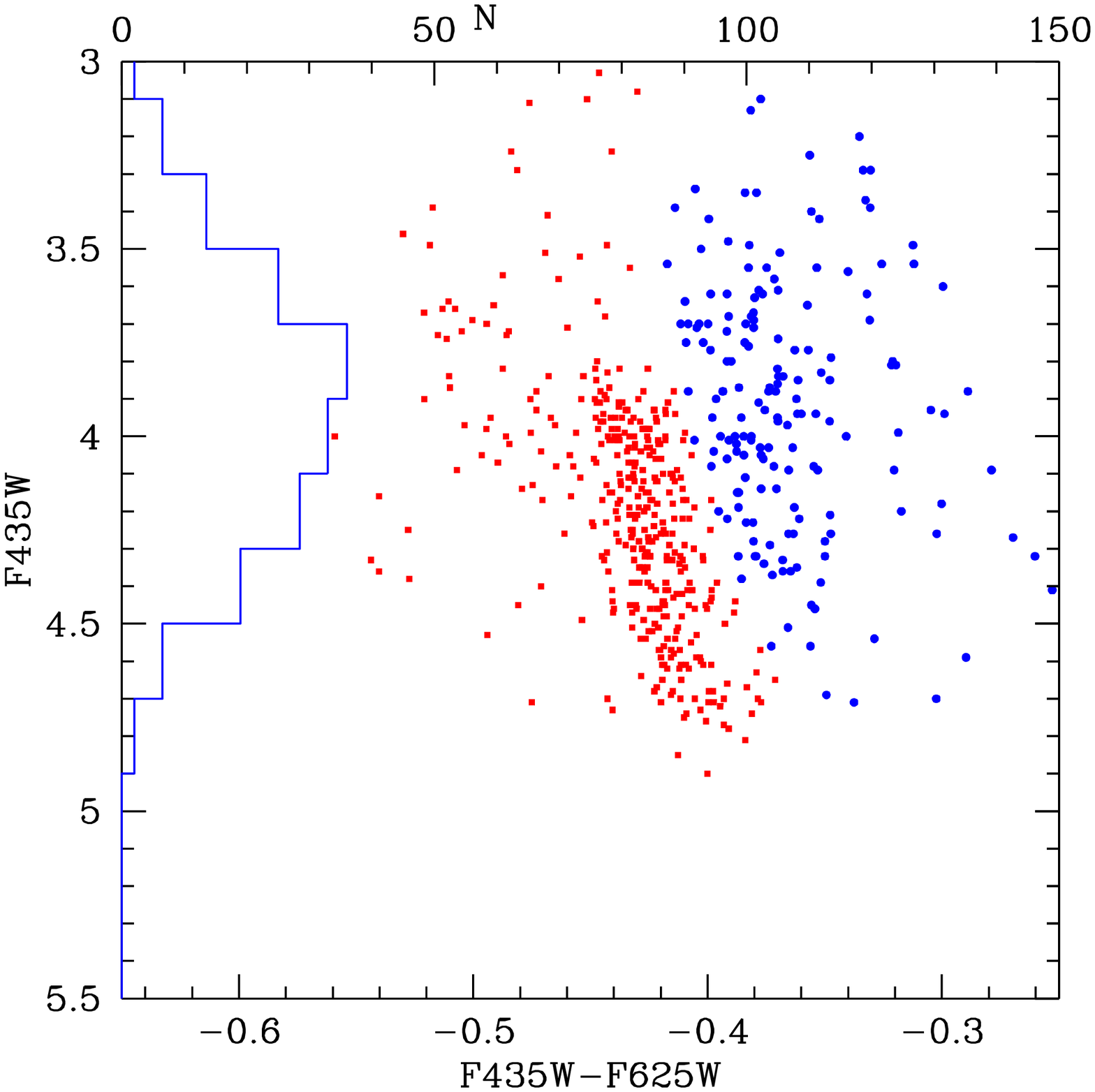}
\includegraphics[width=4.cm]{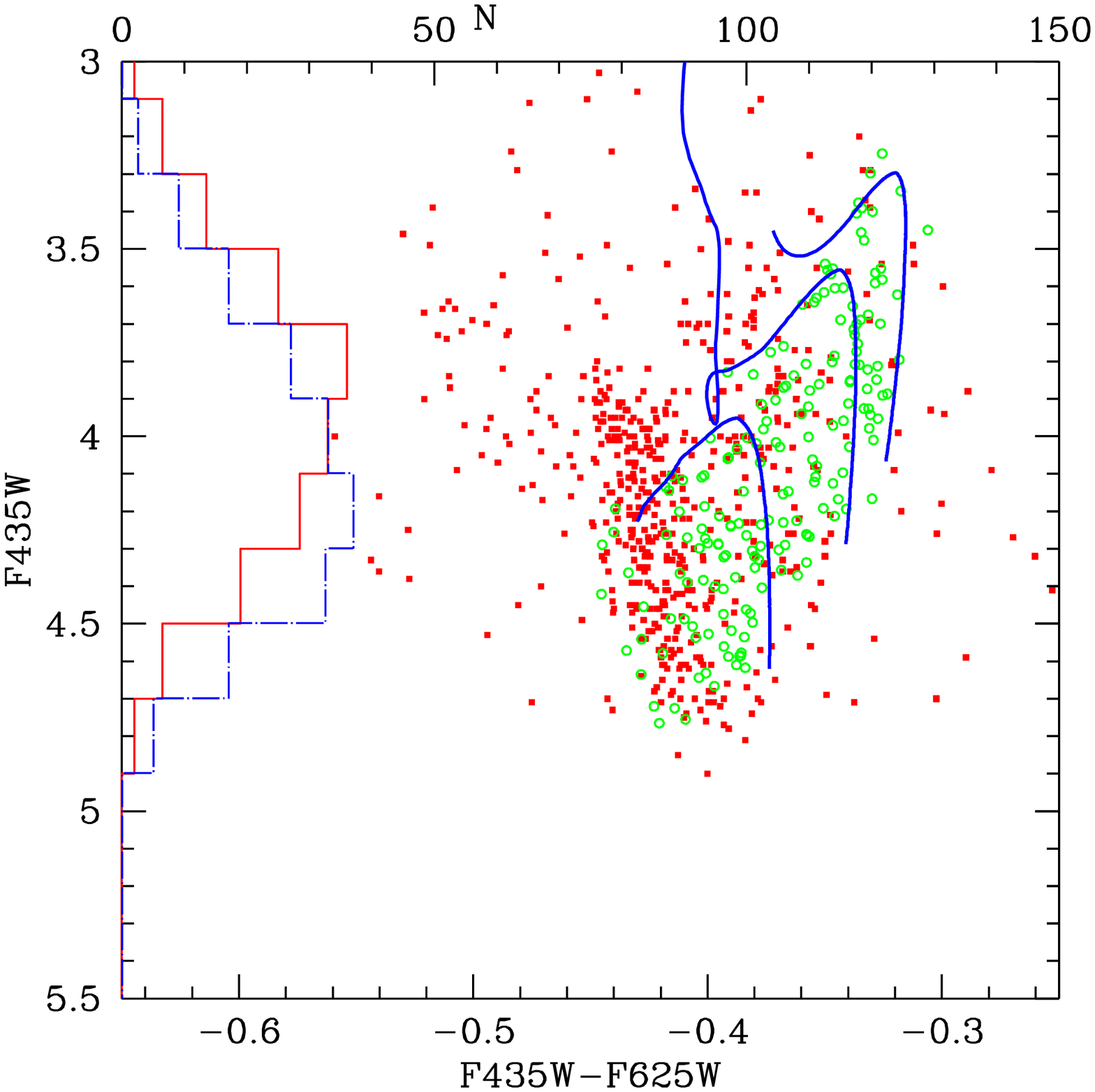}
\includegraphics[width=4.cm]{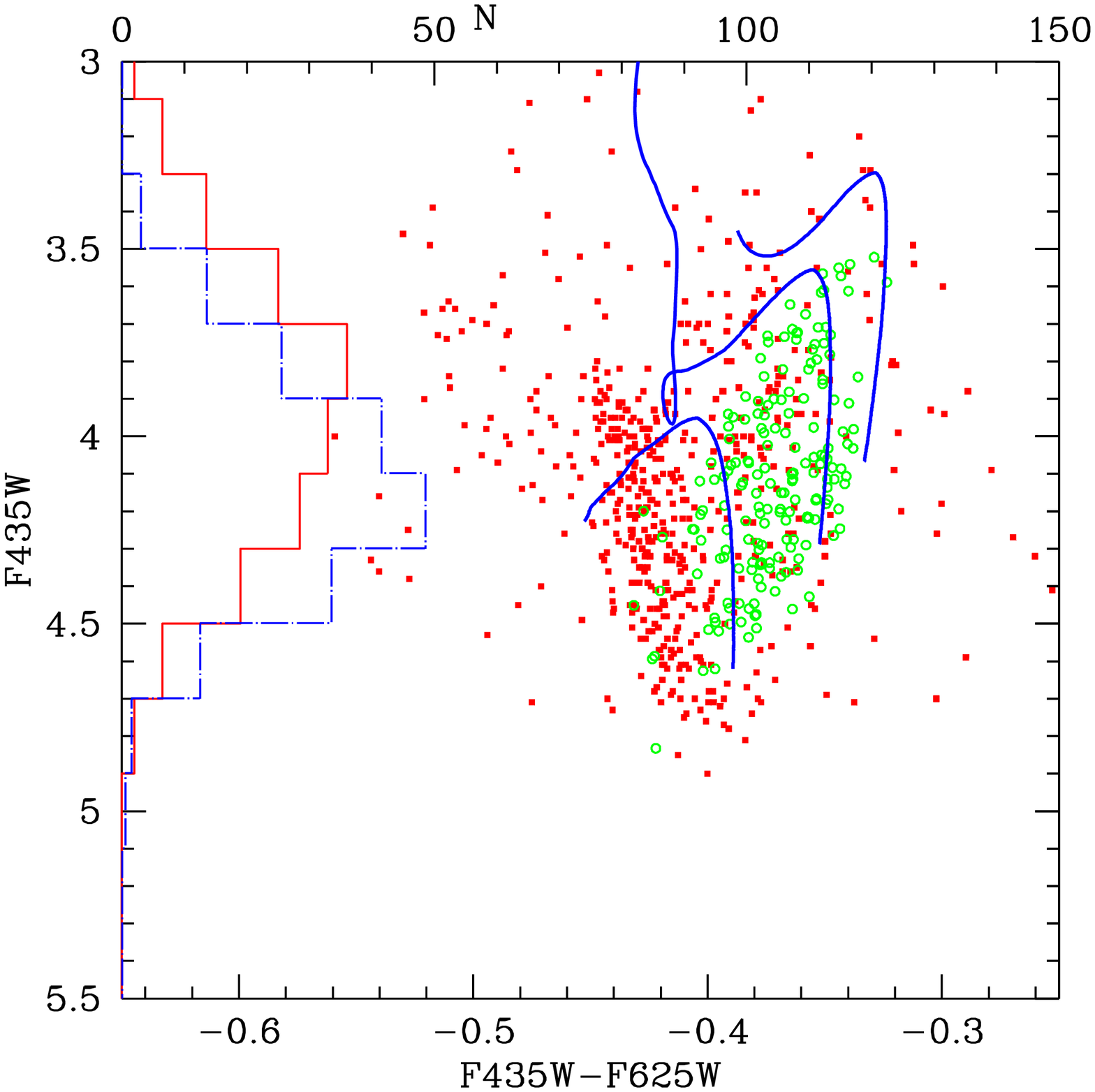}
\includegraphics[width=4.cm]{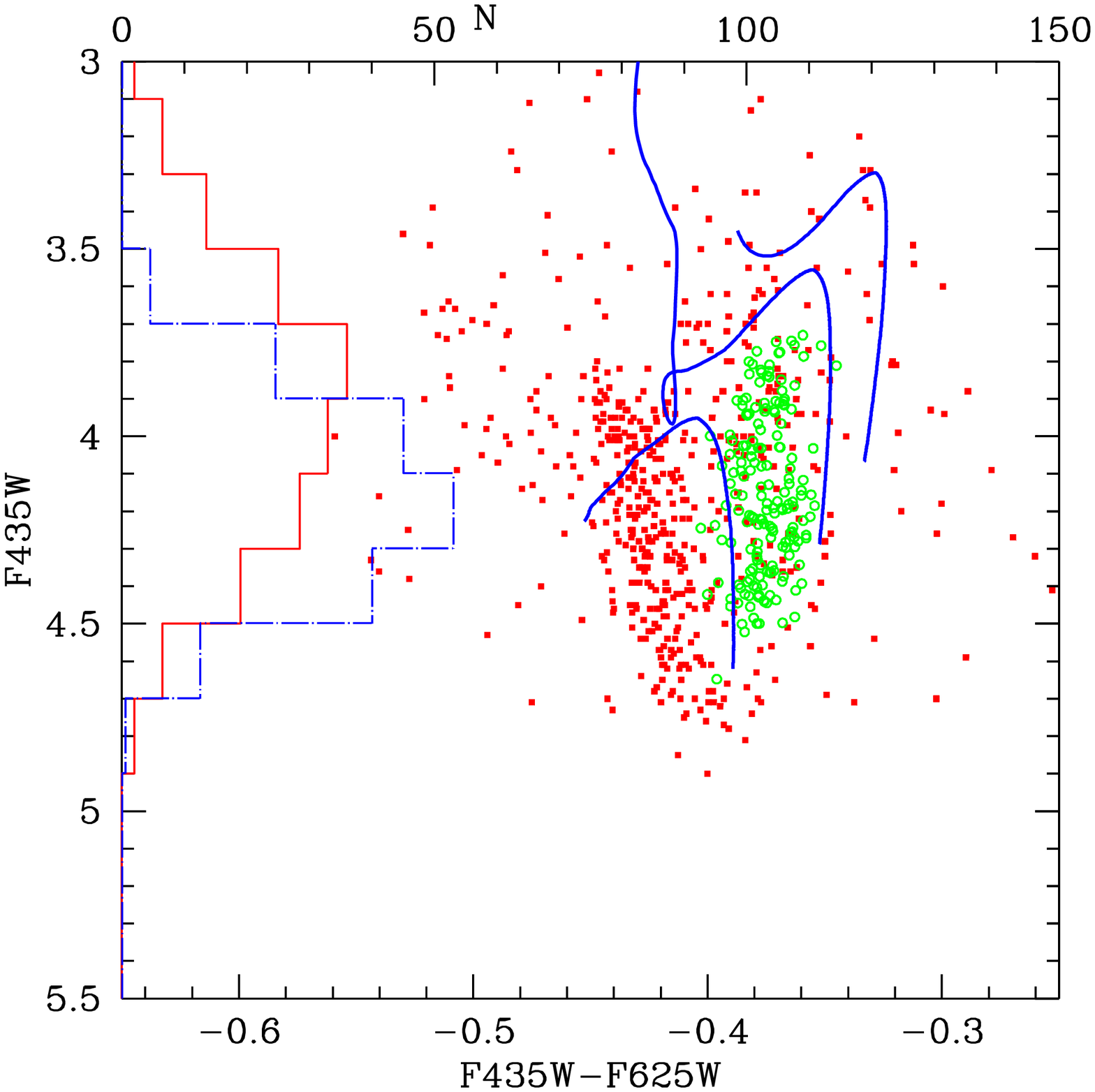}
\caption{The left panel shows again the blue hook, with the identification of the less extreme
population as (blue) dots. On the left we show the number vs. magnitude histogram of the less extreme
sample only. The other panels, from left to right, show the comparison with simulations 
for helium fixed at  Y=0.45, for a mass dispersion of 0.01\msun\, 0.004\msun and 0.002\msun.
Tracks for Y=0.45 and M=0.47, 0.48 and 0.49\msun\ from left to right are overplotted. 
}
      \label{f3}%
\end{figure*}
  
%
  \section{The progenitors of the blue hook stars: non canonical extra--mixing in red giants 
  of the blue MS population?  \label{sec3}}
 
  Here we examine the paths to these peculiar objects that could explain their
  distribution as an evolutionary sequence.\\  
  Notice, first of all, that \ocen\ ``blue MS" has a metal content larger than the
  one of the redder sequence \citep{piotto2005}, so that its location can
  be understood only in terms of an enhanced helium content, of about
  38\%-40\% \citep{norris2004}. Such a large and uniform helium content is in itself
  very difficult to be explained in terms of chemical enrichment \citep[see, e.g.][]{dercole2008},
  but we take this evidence at face value and consider the constraints it poses on the evolution
  of such helium rich stars. Even if the stars populating the blue MS do not differ 
  significantly in age from the rest of the cluster stars\footnote{This can be reasonably assumed, 
  as the only plausible progenitors of the extreme helium rich population are either
  the massive fast rotating stars \citep{decressin2007}, evolving in a few million years, or the
  super--asymptotic giant branch (AGB) stars \citep{pumo2008}, evolving in 30--40 million years.}, 
  they evolve more rapidly, so that along the red
  giant branch we find masses {\it much smaller} than the typical turnoff mass assumed to populate the
  standard--helium main sequence. Using the simple linear approximation given in \cite{dc2008},
  $\delta$M$_{RG}$/$\delta$Y$\sim$--1.3\msun, an increase in helium content from
the primordial value Y = 0.24 to the very high value Y = 0.40
decreases the evolving mass by $\sim$0.18\msun. Although the modalities of mass loss along the RGB
are far from being clear, it is difficult to believe in a strong dependence of mass loss on the
helium abundance, so that we may expect that the giants in a GC lose about the same amount of
mass, irrespectively from their helium content. 
Therefore, it is unavoidable that the progeny of the blue MS stars populates the bluest
parts of the HB, those corresponding to the smallest total masses. In order for a Y=0.24 standard star
to become a hot HB star, we must invoke huge unusual mass loss along the RGB (say, $\sim$
0.15\msun\ larger than those populating the red HB). While this is not
impossible, it is highly unplausible that it concerns a great fraction of the blue hook stars.
On the other hand, being the blue hook stars the HB stars having the smallest total mass, the majority of
the very high helium population must end up there (a fraction can avoid the ignition of the
helium flash and go directly to the helium white dwarf cooling phase).
  {\it This reasoning implies that there must be a direct relation between the helium rich MS stars
  and the blue hook stars}. In addition, the interpretation of the blue hook sequences 
  in terms of even more enhanced
  helium (Y$\sim$0.6--0.8) {\it requires additional mixing in the progenitors 
  of these stars or at the helium flash}.
  Notice that we can not hypothesize that the progenitors of blue hook stars are main sequence stars having
  Y$\sim$0.8 at their birth, for two main reasons: 1) for a reasonable cluster age, and for mass loss
  rate not dramatically smaller than the mass loss of normal--helium populations, the evolving mass
  of Y$\sim$0.8 would not ignite helium, and evolve directly as helium white dwarf; 2) there is no observational
  signal that such stars exist in the cluster.
  
  An attractive explanation for the presence of very helium rich, very hot
  blue hook stars is given by the hypothesis of ``flash mixing" during a late
  central helium flash occurring on the ``peeled--off" red giant \citep{dcruz1996} or even 
  along the white dwarf cooling sequence \citep{brown2001}.
  The more or less abundant mixing of material rich in carbon and helium
  with the (weak) hydrogen shell and the surface layers gives rise, after an
  episode of violent H--burning, to a HB--like structure with an atmosphere
  rich in helium and carbon, as observed. Besides, the luminosity and temperature
  of an object of this kind may be quite different from those of a star lying
  at the extreme of canonical HBs. In particular, they can be quite hotter
  than such stars, in accord with their subluminous position.
  This hypothesis may well be the correct explanation for a few blue hook stars, but certainly 
  can not explain its majority. In fact the remnant hydrogen
  mass on the star suffering a flash after the red giant tip is really very low 
  ---from 8 to 14$\times 10^{-4}$\msun\ in the model calculations by
  \citet{dcruz1996}, see also \citet{miller2008}--- compared to the envelope values of our models 
  (from $\sim$2 to 50$\times 10^{-3}$\msun in the simulations shown in Fig. \ref{f2}).
  In addition, the neat sequences observed in the blue hook seem to derive from a very
  well defined evolutionary phase, not from the erratic outcome of a complex
  process like nuclear burning during a mixing episode. 
  We leave this question open to further computations, and conclude by asking whether 
  is it possible to invoke a more prolonged phase of mixing to raise the envelope helium content
  in all or most of the blue MS stars.

Before the recognition that the GC chemical anomalies are present also at the surface of
scarcely evolved stars \citep{gratton2001,rc02},     
a ``non canonical extra mixing"\footnote{``Canonical" extra--mixing
is defined as the physical process needed to explain the penetration of convection into
the outer part of the H--shell, where the CN branch of the CNO--cycle
is operating, in order to explain the decline of carbon abundance, a strong reduction of the
  $^{12}$C/$^{13}$C isotopic ratio in the upper red giant branch, and the 
decrease in surface Li and $^3$He abundances with increased $L$ \citep{pilach1993,grea00},
  both in field stars and in cluster members  \citep{sneden1991}.}, penetrating deep into the
H--shell, has been hypothesized by \cite{dd90} to explain
the apparent increase in O deficiency and Na abundance in upper giant branch
stars, and even in Al, under certain assumptions (\citealt{lh95,dt00,dw01}). 
Although the multiple population models is now generally accepted,
thanks also to the photometric evidence, deep mixing might 
still play a role in the evolution: \cite{dv03} have proposed that in some upper 
RGB stars canonical extra mixing may be switched to its enhanced mode with much 
faster and somewhat deeper mixing, which could be driven 
by differential rotation of the stellar radiative zones, this latter 
caused, e.g., by the spinning up of close binary members as a result of tidal 
synchronization (\citealt{dea06}).
It is commonly accepted that ``canonical" and ``non--canonical" extra--mixing    
occur at the ``red giant bump" (\citealt{zea99,rea03}), where the H-burning shell, advancing in mass, 
erases the chemical composition 
discontinuity left behind by the bottom of convective envelope at the end of 
the first dredge-up (\citealt{grea00,sh03}). The presence of the red giant bump in
the luminosity function of GCs can be taken as an 
experimental indication that the``standard" models of red giants actually make 
a good job in describing the evolution below the bump luminosity, 
and that no mechanisms of extra 
mixing in low-mass stars are acting, on the lower 
RGB, otherwhise the discontinuity left by convection would be smeared out and 
the bump would not be produced. It is generally assumed that the H--shell 
is shielded against mixing by the gradient in the mean molecular weight associated with the 
composition discontinuity (\citealt{sm79,chea98,dv03}) or, in any case, that it 
operates very slowly (\citealt{chaname,palacios2006}). 
  However, D'Antona \& Ventura (2007) have investigated the evolution of red giant
  stars in the hypothesis that such ``non canonical extra mixing'' is
  efficient only for those cluster members which are most helium rich. The
  practically complete disappearance of the molecular weight
  discontinuity at the end of the first dredge--up (that is also the reason for the
  presence of the red giant ``bump"), is the main hint that such an
  assumption is not unreasonable, and D'Antona \& Ventura (2007) suggest 
  that this deep mixing could well occur early during the red giant evolution, well
  before the bump location. 
  This in situ deep mixing in high--helium red giants evolving 
  today in GCs is able to explain the extremeley low oxygen abundances found in luminous red
  giants in clusters like M13 and NGC~2808. \cite{dv2007} notice that the surface helium
  increases due to this extra--mixing. Although in their models this increase is relatively small,
  an earlier and faster deep mixing can provide stronger enhancements \citep{sweigart1997, weiss2000}.
  In particular, \cite{weiss2000} models including a strong penetration of the mixing, and a large 
  diffusion coefficient show a helium increase 
  in the envelope by $\delta$Y$_{\rm env} \simeq$0.3, close to our
  requirements. These authors discard their models with $\delta$Y$_{\rm env} \geq $0.1,
  noticing that the concomitant increase in sodium and depletion in oxygen would push these
  abundances out of the range observed in the anomalous stars of GCs. 
  Nevertheless, oxygen abundances as low as 
  [O/Fe]$\sim$--1.2 are still acceptable for the extreme anomalies, and realistic models {\it would not}
  show the sodium increase, that is due to mixing through the region in which the $^{22}$Ne+p reaction 
  is active in the giant
  interior. \cite{dv2007} have shown that deep mixing does not alter the sodium abundance, if the 
  evolving giants are formed from the {\it $^{22}$Ne--poor} gas ejected by the massive AGBs subject to HBB. 
  Therefore, the models with very deep mixing can not be excluded on the basis of the extreme sodium
  anomalies predicted by models that do not start with the correct chemistry of the HBB processed
  gas.
  
  Another approach is the ``flash assisted" mixing
  proposed by \cite{fuji1999}. This model shows that,
  if hydrogen is carried down into the helium core by some extra mixing mechanism, a
hydrogen-burning shell flash is ignited. The flash forces the formation of a convective shell 
whose outer edge extends into hydrogen-rich layers, bringing in fresh hydrogen to fuel the flash 
further, and 
during the decay phase of the flash the nuclear products are dredged up by surface convection, 
which becomes deeper in mass than during the quiescent phases. Also in these models, the
surface helium enrichment may become very large, and the surface abundances will resemble
the extreme anomalies found in GCs \citep{aikawa2001}.
  
  We propose than that deep mixing acts only in the very high helium giants, 
  progeny of the blue main sequence. Even starting from Y$\sim$0.38--0.40, these stars
  can reach the required very high helium in the envelope, prior to the helium flash and to 
  their subsequent evolution, that consequently occurs along the blue hook. 
  Other (unknown) physical inputs may cause a bimodal deep mixing, with different 
  final helium content and the presence of two sides of the blue hook.
  The double character of the blue hook in \ocen\ seems in fact to indicate that there are actually two
  kinds of helium rich progenitors of the blue hook, the most dominant ending with Y$\sim$0.8
  in the envelope, the latter one ending with a bit lower helium. Further observations and theoretical
  work are needed to understand this result.
  Model computation for this very deep extramixing is needed to understand 
  whether this is a viable solution to our problem.
  
\section{Acknowledgments} 
This work has been supported through PRIN MIUR 2007 
``Multiple stellar populations in globular clusters: census, characterization and
origin". We thank Giampaolo Piotto, Alvio Renzini and Sandro Villanova
for stimulating discussions, and J. Anderson, L. Bedin, 
and A. Milone for sharing their data with us. We finally thank the anonymous referee,
whose comments helped us to clarify several points of the discussion.

\label{lastpage}

\end{document}